\documentclass[manuscript]{aastex63}

\usepackage{color}
\usepackage{CJK}
\newcommand{\cntext}[1]{\begin{CJK}{UTF8}{gbsn}#1\end{CJK}\kern-1ex}

\begin{document}
\title{Double-power-law feature of energetic particles accelerated at coronal shocks}

\correspondingauthor{Xiangliang Kong}
\email{kongx@sdu.edu.cn}

\author[0000-0002-1576-4033]{Feiyu Yu}
\affiliation{School of Space Science and Physics, Institute of Space Sciences, Institute of Frontier and Interdisciplinary Science, Shandong University, Shandong, People's Republic of China}

\author[0000-0003-1034-5857]{Xiangliang Kong}
\affiliation{School of Space Science and Physics, Institute of Space Sciences, Institute of Frontier and Interdisciplinary Science, Shandong University, Shandong, People's Republic of China}

\author[0000-0003-4315-3755]{Fan Guo}
\affiliation{Los Alamos National Laboratory, Los Alamos, NM 87545, USA}
\affiliation{New Mexico Consortium, 4200 West Jemez Rd, Los Alamos, NM 87544, USA}

\author{Wenlong Liu}
\affiliation{School of Space Science and Physics, Institute of Space Sciences, Institute of Frontier and Interdisciplinary Science, Shandong University, Shandong, People's Republic of China}

\author{Zelong Jiang}
\affiliation{School of Space Science and Physics, Institute of Space Sciences, Institute of Frontier and Interdisciplinary Science, Shandong University, Shandong, People's Republic of China}

\author[0000-0001-6449-8838]{Yao Chen}
\affiliation{School of Space Science and Physics, Institute of Space Sciences, Institute of Frontier and Interdisciplinary Science, Shandong University, Shandong, People's Republic of China}

\author[0000-0002-0850-4233]{Joe Giacalone}
\affiliation{Department of Planetary Sciences, University of Arizona, Tucson, AZ 85721, USA}

\begin{abstract}
Recent observations have shown that in many large solar energetic particle (SEP) events the event-integrated differential spectra resemble double power laws.
We perform numerical modeling of particle acceleration at coronal shocks propagating through a streamer-like magnetic field by solving the Parker transport equation, including protons and heavier ions.
We find that for all ion species the energy spectra integrated over the simulation domain can be described by a double power law, and the break energy depends on the ion charge-to-mass ratio as $E_B \sim (Q/A)^\alpha$, with $\alpha$ varying from 0.16 to 1.2 by considering different turbulence spectral indices.
We suggest that the double power law distribution may emerge as a result of the superposition of energetic particles from different source regions where the acceleration rates differ significantly due to particle diffusion. 
The diffusion and mixing of energetic particles could also provide an explanation for the increase of Fe/O at high energies as observed in some SEP events.
Although further mixing processes may occur, our simulations indicate that either power-law break or rollover can occur near the Sun and predict that the spectral forms vary significantly along the shock front, which may be examined by upcoming near-Sun SEP measurements from Parker Solar Probe and Solar Orbiter.

\end{abstract}



\section{Introduction} \label{sec:intro}

During explosive solar activities, a large number of charged particles are accelerated to high energies near the Sun. As a consequence, the radiation level of energetic particles in the near-Earth space can be significantly enhanced, known as solar energetic particle (SEP) events \citep{reames99,desai16}.
Large SEP events can pose serious threats to satellites and astronauts in space, therefore they are of particular importance to space weather.
In these events, energetic particles are mainly accelerated by the shocks driven by fast coronal mass ejections (CMEs).
{\it In situ} observations at 1 AU have shown that the fluence spectra (integrated over the individual events) of energetic protons and heavy ions often exhibit double power laws and the spectral break energy depends on the ion charge-to-mass ratio as $E_B \sim (Q/A)^\alpha$ \citep[e.g.,][]{cohen05,mewaldt05,mewaldt12,desai16a,desai16b,cohen18}.

\begin{figure}
\centering
\includegraphics[width=0.95\linewidth]{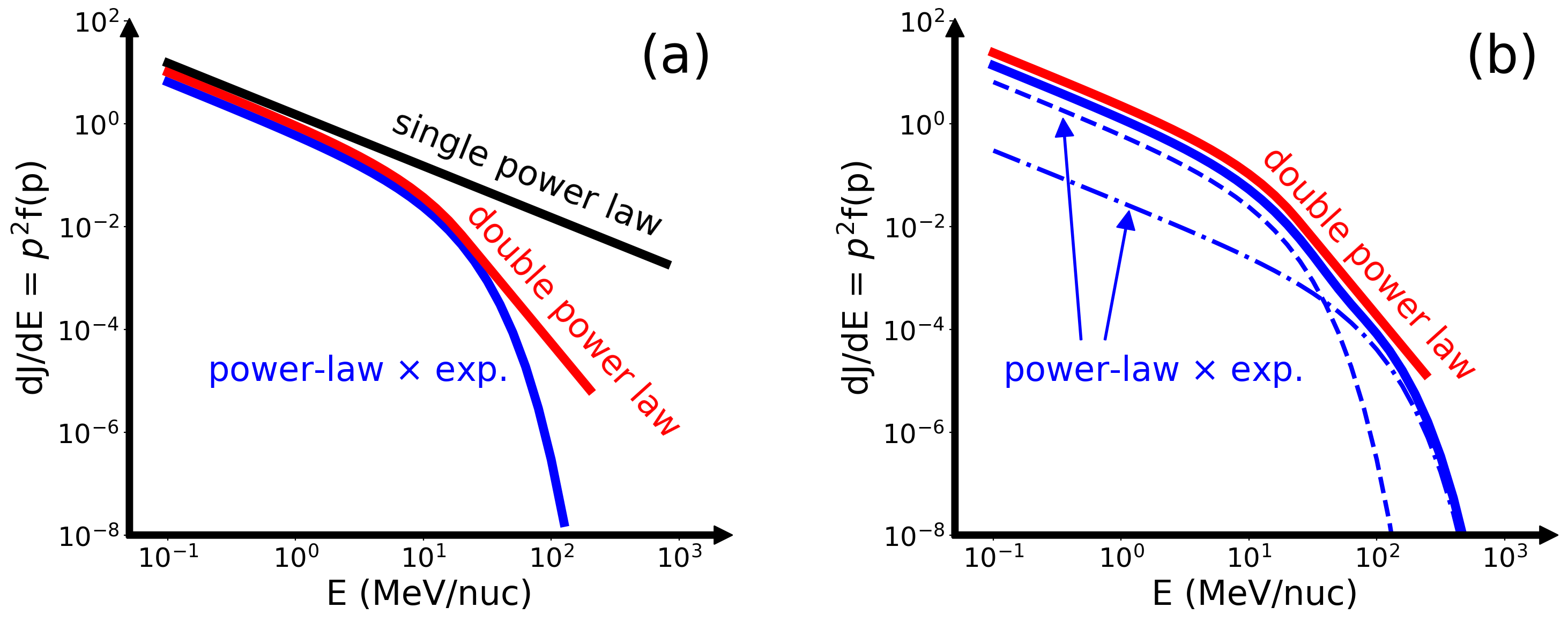}
\caption{(a) DSA theory predicts a single power law for particles accelerated by a 1D planar shock, while the particle energy spectra often exhibit an exponential rollover at high energies \citep{ellison85}.
The fluence spectra in some large SEP events are better fitted by double power-laws \citep{band93}.
(b) A superposition scenario to explain the formation mechanism of the double power-law energy spectrum.
The blue solid line is obtained by the superposition of two energy spectra of ``power-law $\times$ exponential'' form but with different break energies ($E_B$ = 10, 50 MeV/nuc). 
The superposed energy spectrum resembles a double power-law below a certain energy ($\sim$200 MeV/nuc). 
}
\label{fig:mixing}
\end{figure}

The double-power-law feature observed in SEP spectra puts critical constraints on the physics affecting the acceleration and transport of energetic particles. However, the formation mechanism remains unclear. While it may emerge as a result of particle acceleration at CME-shocks close to the Sun \citep{li05a,li05b,li09,tylka06a,schwadron15,kong19}, it may also be due to transport effects in the interplanetary space \citep{li15,zhao16,zhao17}.
The species-independence of spectral indices and the systematic $Q/A$ dependence of $E_B$, as well as the $\alpha$ values, may suggest that the double power-law SEP energy spectra occur due to diffusive shock acceleration (DSA) near the Sun rather than interplanetary scattering \citep{desai16b}.
The DSA theory predicts a steady-state solution of a power-law energy spectrum with the spectral index only depending on the compression ratio for a 1D planar shock.
The shock-accelerated particles can exhibit an exponential rollover at high energies due to various effects including the finite shock size, limited acceleration time, shock geometry, and adiabatic cooling \citep{ellison85,zank00}.
However, the standard DSA theory could not explain how the second power-law above the break is produced (see Figure \ref{fig:mixing}(a)).
\citet{li05a} introduced a loss term in the Parker transport equation and showed that a broken power-law can form if the escaping effect suddenly kicks in above a certain energy.
\citet{schwadron15} studied particle acceleration at size-limited coronal shocks based on analytic solutions of the transport equation and found that double power-law distributions can form due to particle diffusion and escape from the acceleration sites.

CME-driven shocks can form near the Sun and accelerate particles to $\sim$GeV in GLE events \citep[e.g.,][]{reames09,gopalswamy12,gopalswamy13}.
When the shock sweeps through the complex magnetic field structures in the corona, the shock geometry varies both along the shock front and with the shock propagation \citep{sandroos09,kozarev15,kong16,kong17,rouillard16,kouloumvakos19}, which strongly affects the rate of particle acceleration \citep{giacalone05a,giacalone05b,tylka05}.
An example of the magnetic structures are streamers, the most prominent quasi-steady structures in the corona and commonly interacting with CME-shocks.
For instance, CME-streamer interactions can be observed in the two GLE events in solar cycle 24 \citep{kong19}. In some SEP events, streamers may play an important role in the acceleration, trapping and release of energetic particles \citep[e.g.,][]{kocharov17,cliver20,kouloumvakos20}.
Recently, \citet{kong17,kong19} numerically modeled the acceleration of protons at a coronal shock and found that the shock-streamer interaction region favors more efficient particle acceleration due to both trapping effect of closed fields and quasi-perpendicular shock geometry. This suggests that the coronal magnetic field configuration can be an important factor for producing large SEP events.
\citet{kong19} also found the emergence of double power-law feature in particle spectra and suggested that it may be a mixture of two distinct populations accelerated in the streamer and open field regions, where the particle acceleration rate differs largely.
As shown in Figure \ref{fig:mixing}(b), the superposition of two particle spectra of ``power-law with exponential rollover" but with different break energies can lift up the high-energy spectrum and results in a second power-law feature below a certain energy.

In this work, we investigate the acceleration of both protons and heavier ions at coronal shocks and focus on the formation mechanism of double power laws.
Our simulations show that the double power law distribution can occur during particle acceleration at near-Sun shocks due to the superposition of energetic particles from different source regions.
This paper is organized as follows. In Section \ref{sec:Numerical Model}, we introduce our numerical model. In Section \ref{sec:Simulation Results}, we present the simulation results including spatial distributions and energy spectra of accelerated particles, and examine the $Q/A$ dependence of break energy.
We explain the mechanism of the double power-law energy spectrum based on the superposition scenario as shown in Figure \ref{fig:mixing}.
We also discuss the increase of Fe/O at high energies as observed in some SEP events based on our simulations.
Conclusions and discussion are given in Section \ref{sec:Conclusions and Discussion}.

\section{Numerical Model} \label{sec:Numerical Model}

Following our previous studies \citep{kong17,kong19}, we explore numerically the acceleration of energetic particles at a coronal shock.
The shock is described in a kinematic approach and simplified as an expanding circular front.
The center of the shock is fixed in the solar equatorial plane at the height of 1.1 $R_\odot$ and the shock propagates outward with a constant speed $V_{sh}$ = 2000 km s$^{-1}$ and a compression ratio $X = 3$. The background magnetic field is given analytically as a partially open streamer-like magnetic field \citep{low86}. The magnetic field in the downstream region is compressed by the shock  and calculated from the ideal MHD induction equation.
Here we consider the configuration when a CME-driven shock sweeps through the streamer magnetic field from its flank (Figure \ref{fig:streamer_5o3}(a)).

We model the acceleration and transport of particles by numerically solving the Parker transport equation \citep{parker65}. It is achieved by integrating stochastic differential equations corresponding to the Fokker–Planck form of the transport equation using a large number of pseudo-particles \citep[e.g.,][]{zhang99,guo10,kong17,li2018,kong2019b}. 
In addition to protons, we also consider different elements, He, O, Mg, and Fe. Their charge-to-mass ratios are taken as $Q/A$ = 2/4, 7/16, 9/24, 12/56, respectively \citep{cohen05,zhao16}.

In the corona, the presence of solar wind turbulence can significantly enhance the trapping of charged particles near the shock front and affect the acceleration and transport of energetic particles.
We consider the turbulence power spectrum in the coronal region as \citep{giacalone99}
\begin{equation}
P(k) = A_0 L_c \sigma^2 B_0^2 \frac{1}{1 + (k L_c)^{\Gamma}},
\end{equation}
where $A_0$ is the normalization constant, $L_c$ is the correlation length, $\sigma^2$ is the turbulence variance, $B_0$ is the background magnetic field, $k$ is the wave number, and $\Gamma$ is the spectral index.

As shown in the Appendix, according to the quasi-linear theory \citep{jokipii66}, we deduce the expression of the parallel diffusion coefficient for different ions as
\begin{equation}
\kappa_{\parallel} = \kappa_{\parallel 0} \\
\frac{\gamma_0}{\gamma} \\
\left( \frac{p}{p_0} \right)^{3 - \Gamma} \\
\left( \frac{Q}{A} \right)^{\Gamma - 2},
\end{equation}
where $p$ is the particle momentum, $\gamma$ is the Lorentz factor, and the subscript $0$ refers to that for the injection energy.
For Kolmogorov turbulence spectrum ($\Gamma = 5/3$), the parallel diffusion coefficient of protons with the initial energy of $100$ keV is taken as $\kappa_{\parallel 0} = 1.4 \times 10^{17}$ cm$^2$ s$^{-1}$.
Since we only focus on the low coronal region, we assume that the turbulence variance and coherence length are independent of distance from the Sun.
In the simulations, we normalize the length  $L_0$ = 1 $R_\odot$ = 7 $\times$ $10^{5}$ km and the velocity $V_0$  = $V_{sh}$ = 2000 km s$^{-1}$. 
So the normalization of the diffusion coefficient is $\kappa _0$ = $L_0 V_0$ = 1.4 $\times$ $10^{19}$ cm$^2$ s$^{-1}$, then $\kappa_{\parallel 0}$ = 0.01 $\kappa_0$.
In addition to Kolmogorov turbulence spectrum, we also consider different turbulence spectra by varying the  spectral indices $\Gamma$ = 1.9, 1.1, and 0.5.
$\kappa_{\parallel 0}$ is taken as 0.01 $\kappa_0$ for $\Gamma$ = 1.9 and 0.001 $\kappa_0$ for $\Gamma$ = 1.1 and 0.5.

Charged particles normally move along magnetic field lines, but the pitch-angle scattering, drift, and field line random walks can cause perpendicular diffusion.
Test-particle calculations show that the perpendicular diffusion coefficient $\kappa_{\perp}$ is approximately a few percent of the parallel diffusion coefficient $\kappa_{\parallel}$ and that $\kappa_{\perp} / \kappa_{\parallel}$  is independent of ion energy \citep{giacalone99}. We take $\kappa_{\perp}$ = 0.01 $\kappa_{\parallel}$ in this work.

As the shock propagates outward, we inject the ions with an initial energy of 100 keV/nuc at a constant rate into upstream of the shock.
For different elements, we inject the same number of pseudo-particles ($\sim$ half of a million) and normalize the density by their coronal abundance.
To improve the statistics of high-energy particle distribution, we use a particle splitting technique \citep[e.g.,][]{kong17,kong19}.
The particle injection energy is approximately equal to what is required to describe the acceleration process in Parker transport equation, i.e., the streaming anisotropy is sufficiently small \citep{giacalone99,Guo2021}.
We only focus on particle acceleration and transport in low coronal region, therefore the simulation is terminated when the shock travels to 4 $R_{\odot}$. A particle will be removed from the simulation if it reaches the $z$-axis or the solar surface.

\section{Simulation Results} \label{sec:Simulation Results}

\begin{figure}
\centering
\includegraphics[width=0.95\linewidth]{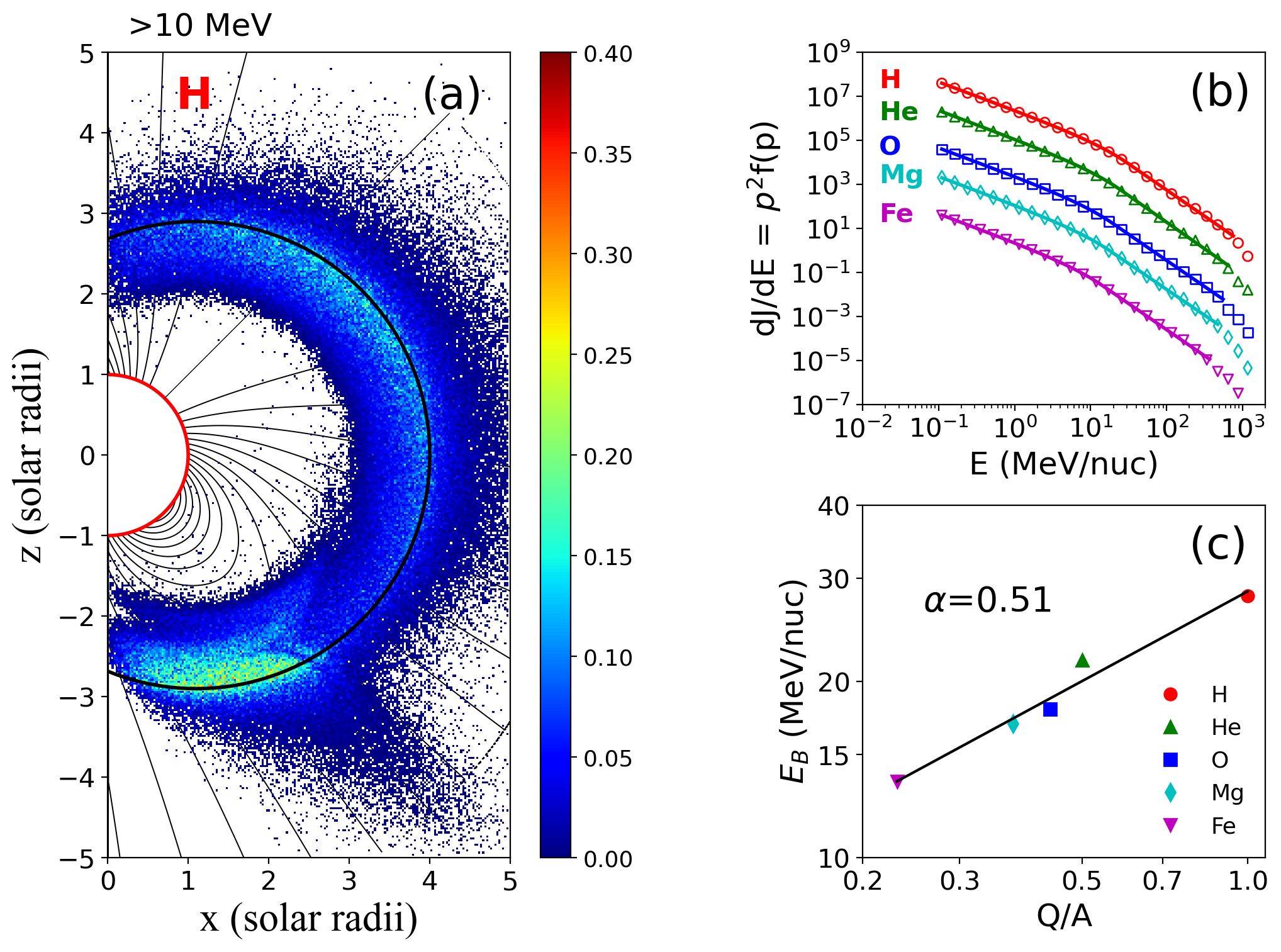}
\caption{(a) Spatial distribution of accelerated protons with energies $>$10 MeV as the shock propagates to 4.0 $R_{\odot}$. 
The shock front is indicated by the black circle.
(b) Energy spectra of different ions (H, He, O, Mg, and Fe) accelerated by the shock in Kolmogorov turbulence spectrum.
The solid curves are the fitting energy spectra using the Band function \citep{band93} with low-energy indices $\gamma_1 \sim 1.3$ and high-energy indices $\gamma_2 \sim 2.4$.
(c) Spectral break energy $E_B$ vs. the ion charge-to-mass ratio Q/A.
The solid line shows the fitting with a power-law function $E_B \sim (Q/A)^{\alpha}$. 
}
\label{fig:streamer_5o3}
\end{figure}

We first investigate the acceleration of different ion species, i.e., H, He, O, Mg, and Fe, in the case of the Kolmogorov turbulence spectrum.
Figure \ref{fig:streamer_5o3}(a) shows the spatial distribution of accelerated protons with energies $>$10 MeV when the shock moves to 4 $R_{\odot}$. 
The upper flank of the shock is defined as the nonstreamer region while the lower flank as the streamer region.
The intensity of high-energy particles in the streamer region is much higher than that in the nonstreamer region.
This indicates that the trapping effect of closed magnetic field when the shock sweeps through a streamer can result in more effective particle acceleration \citep{kong17,kong19}.
For heavy ions, the spatial distributions show consistent results (see below).

Figure \ref{fig:streamer_5o3}(b) shows the energy spectra of different ion species integrated over the whole simulation domain, 
which all exhibit double power laws at energies less than 300$-$500 MeV/nuc.
We fit the particle spectra using the Band function \citep{band93},
\begin{equation}
\frac{dJ}{dE} = \left\{
\begin{array}{rcl}
& C E^{-\gamma_1}\exp(-\frac{E}{E_B}) & for {E \leq (\gamma_2 - \gamma_1)E_B}, \\
& C E^{-\gamma_2}[(\gamma_2 - \gamma_1)E_B]^{\gamma_2 - \gamma_1}\exp(\gamma_1 - \gamma_2) & for {E \geq (\gamma_2 - \gamma_1)E_B},
\end{array} \right.
\end{equation}
where $C$ is the normalization constant, $\gamma_1$ and $\gamma_2$ are the power-law indices for low-energy and high-energy particles, and $E_B$ is the break energy.
The fitting parameters for H, He, O, Mg, and Fe are $\gamma_1$ = 1.29, 1.30, 1.29, 1.29, 1.30;  $\gamma_2$ = 2.37, 2.42, 2.37, 2.35, 2.37; $E_B$ = 28, 22, 18, 17, 13 MeV/nuc, respectively.
In our model, the shock compression ratio is a constant $X = 3$.
The DSA theory predicts a particle spectrum $f(p) \propto p^{-3X/(X-1)} = p^{-4.5}$, corresponding to the differential intensity $dJ/dE = p^2 f(p) \propto E^{-1.25}$.
Therefore, the low-energy spectral indices of different ion species are close to the theoretical prediction.
We note that an additional rollover appears at several hundred MeV/nuc as observed in large SEP events \citep{bruno18,bruno19}.
For heavier ions (smaller $Q/A$), their break energies $E_B$ are lower.
Figure \ref{fig:streamer_5o3}(c) shows $E_B$ as a function of  $Q/A$. 
By fitting the distribution with a power law function $E_B \sim (Q/A)^{\alpha}$, we find $\alpha=0.51$.
\citet{cohen05} suggested that the spectral breaks of different species should occur at the same value of diffusion coefficient. Then it can be inferred that $\alpha = 2(2-\Gamma)/(3-\Gamma)$  by assuming the turbulence spectrum in the form of $P(k) \sim k^{- \Gamma}$ \citep{droge94,cohen05}.
For the Kolmogorov turbulence spectrum, $\Gamma = 5/3$, it gives $\alpha = 0.5$, agreeing with our simulation results.

\begin{figure}
\centering
\includegraphics[width=0.8\linewidth]{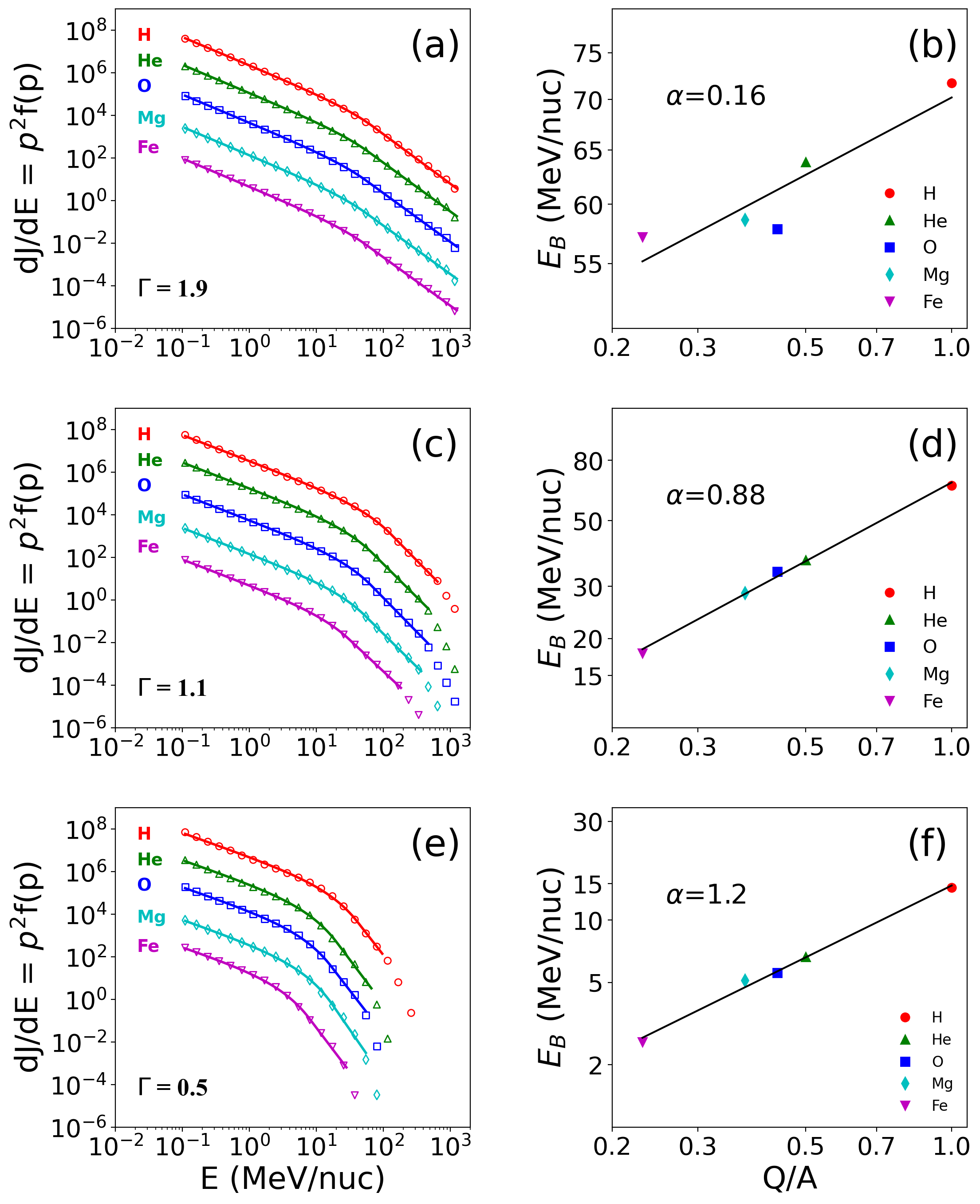}
\caption{(a), (c) and (e): Energy spectra of accelerated particles, with the spectral indices of magnetic turbulence being 1.9, 1.1, and 0.5, respectively.
The solid curves are the fitting spectra using the Band function.
(b), (d), and (f): Spectral break energy $E_B$ vs. the ion charge-to-mass ratio Q/A.
}
\label{fig:different}
\end{figure}

The turbulent magnetic field in the corona and solar wind can affect the acceleration and transport of energetic charged particles.
In Figure \ref{fig:different} we show the simulation results for three cases with different turbulence spectral indices, $\Gamma$ = 1.9, 1.1, and 0.5. The energy spectra of different species integrated over the whole simulation domain are all approximately double power laws and we fit the spectra using the Band function (Equation 3). 
When $\Gamma = $ 1.9, the break energies $E_B$ of different ion species are 72, 64, 58, 59, and 57 MeV/nuc;
when $\Gamma = 1.1$, $E_B$ = 66, 37, 34, 29, and 18 MeV/nuc;
when $\Gamma = 0.5$, $E_B$ = 14, 6.7, 5.6, 5.1, and 2.6 MeV/nuc, respectively.
We also examine the correlation between $E_B$ and Q/A by using a power-law function, $E_B \sim (Q/A)^{\alpha}$, and find that $\alpha$ is in the range of 0.16$-$1.2.
Our simulation results agree well with the ``equal diffusion coefficient'' condition proposed by \citet{cohen05}.
Note that previous numerical simulations of acceleration of heavy ions at coronal shocks have also shown that the rollover energies depend on $Q/A$ \citep[e.g.,][]{sandroos07,battarbee11}, although they did not obtain any double-power-law feature.
This indicates that the $Q/A$ dependence of spectral break/rollover is a common result of DSA.

\begin{figure}
\centering
\includegraphics[width=0.95\linewidth]{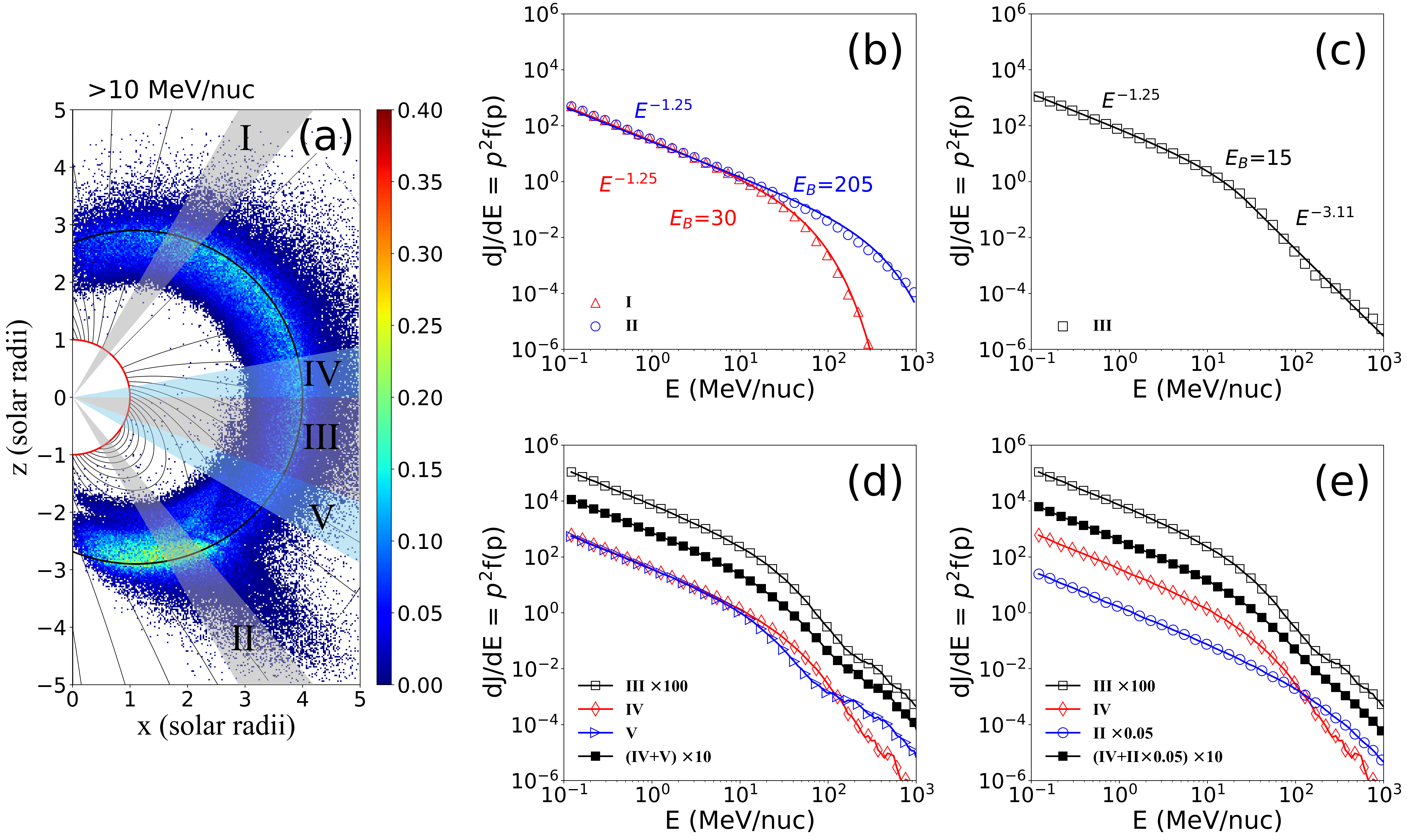}
\caption{(a): Same to Figure \ref{fig:streamer_5o3}(a), spatial distribution of accelerated protons with energies $>$10 MeV as the shock propagates to 4.0 $R_{\odot}$.
(b) and (c): Energy spectra of accelerated protons in regions I and II, and region III, respectively.
Solid lines are the fitting spectra using the exponential rollover function in regions I and II, and the Band function in region III.
(d): Energy spectra in regions IV and V, and the superposed energy spectrum of these two regions.
(e): Energy spectra in regions IV and II, and the superposed energy spectrum of region IV and 0.05 $\times$ region II.
}
\label{fig:double}
\end{figure}

In our previous work, \citet{kong19} found that the proton spectra integrated over the whole simulation domain display a double power law feature and suggested that it may be due to a mixture of two distinct populations accelerated in the streamer and nonstreamer regions.
As shown in Figure \ref{fig:mixing}(b), here we refer it as a superposition scenario. 
We now further explore the formation mechanism of the double power law distribution by examining particle spectra in localized regions along the shock front.
As shown in Figure \ref{fig:double}(a), we first select three localized regions, the nonstreamer region, the streamer region, and the transition region, referred to as regions I, II, and III, respectively.
The particle spectra in regions I and II drop rapidly at high energies and are better fitted with the form of ``power-law $\times$ exponential rollover", as suggested by \citet{ellison85}. When we fit the spectra, the low-energy spectral index is fixed as $\gamma_1$ = 1.25, and we obtain the rollover energies $E_B$ = 30 and 205 MeV/nuc, respectively. The spectral rollover energy in region II is much higher than that in region I, which also shows that particles can be more efficiently accelerated in the streamer region. 
In region III, the high-energy interval of the spectrum apparently displays a power-law, and the spectrum can be well fitted with the Band function.
The Band parameters are $\gamma_1$ = 1.25, $\gamma_2$ = 3.11, and $E_B$ = 15 MeV/nuc.
We also find that both the rollover energy in region I and the break energy in region III scale with Q/A, with the index $\alpha$ similar to that found in the energy spectra integrated over the whole simulation domain as shown above.

To understand the formation of double power law in region III, we select two other regions nearby region III, referred to as regions IV and V. As shown in Figure \ref{fig:double}(d), the particle spectra in the two regions are nearly identical below 10 MeV, but differ substantially at high energies. 
In region V, which is closer to the streamer axis, the spectrum above 100 MeV clearly has a bump and is much flatter than that in region IV.
If we compare the high-energy flux of the spectra from regions II, V, III, IV, to I, it exhibits a decreasing trend according to the distance from the streamer axis. 
Combining the spatial distribution of accelerated protons, we suggest that $>$100 MeV particles in regions III, IV, and V mainly originate from the streamer region (region II) due to the transport effect, and the double power law distribution in region III may occur as a result of the superposition of these high-energy particles with the locally accelerated relatively lower-energy particles.
In Figure \ref{fig:double}(d), we show that the combination of spectra in regions IV and V can give rise to a spectrum comparable to that in region III.
In addition, as shown in Figure \ref{fig:double}(e), if we assume a small fraction (5\%) of high-energy particles escape from region II and mix with the particles in region IV, the compound spectra can produce a double power law, which also looks similar to that in region III.


According to the DSA theory, heavier ions are less efficiently accelerated. If the shock-accelerated particles can be typically described by the ``power-law $\times$ exponential rollover'' spectral form \citep{ellison85}, the Fe/O ratio is expected to fall rapidly with increasing energy.
However, in some large SEP events, such as the event on 2002 August 24, at high energies above $\sim$10 MeV/nuc, Fe/O increases with increasing energy \citep{tylka05}.
The mechanism of the variability in the energy-dependent behaviour of Fe/O remains unsolved and most proposed ideas take into account energetic particles associated with solar flares, either direct contributions or reacceleration by CME-shocks (see Section 2.6 in \citet{desai16} for a review). Here we suggest a different mechanism that may provide an explanation for this phenomenon.

\begin{figure}
\centering
\includegraphics[width=0.95\linewidth]{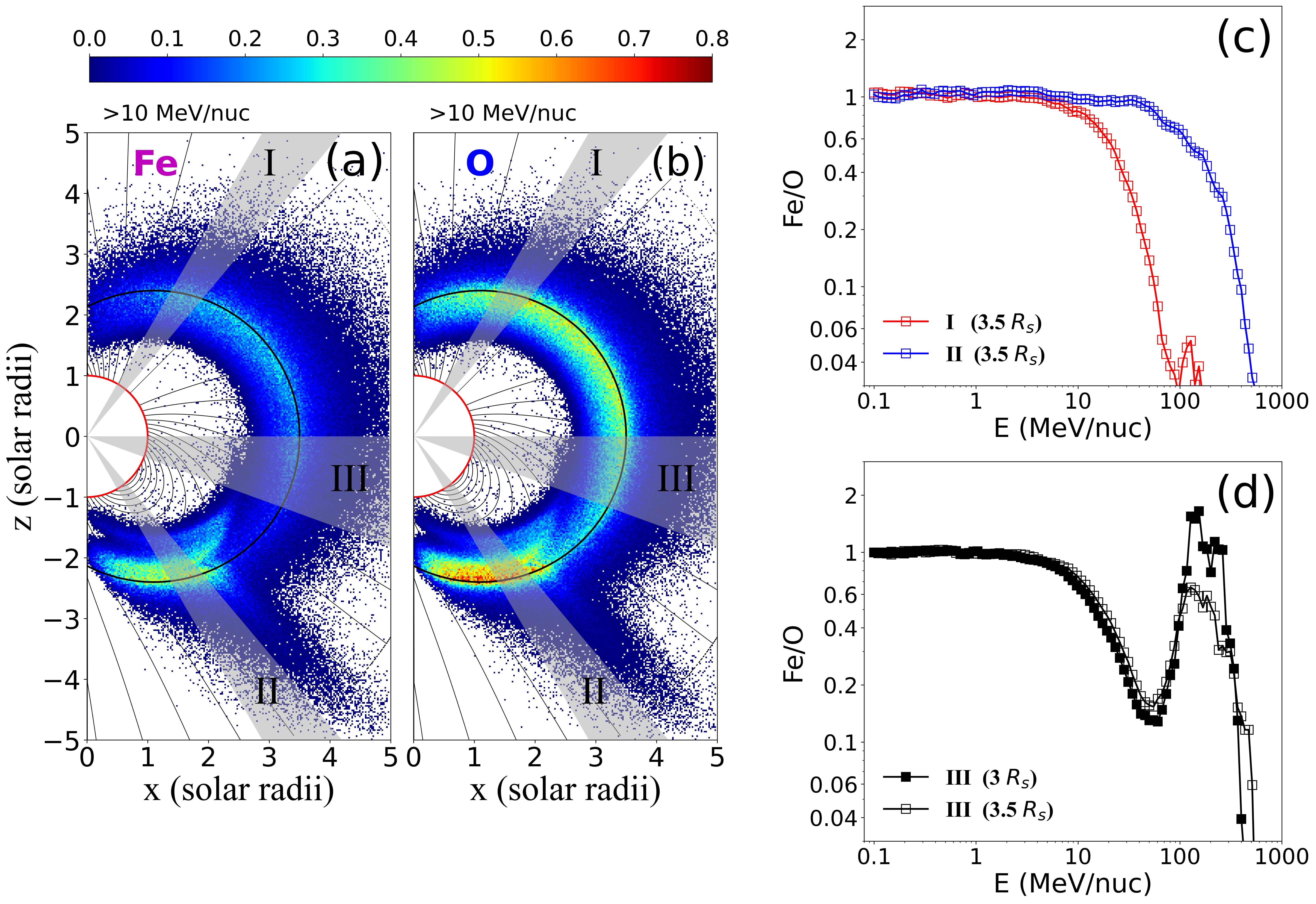}
\caption{(a) and (b): Spatial distributions of Fe and O with energies $>$10 MeV/nuc as the shock propagates to 3.5 $R_{\odot}$.
(c) and (d): Variations of Fe/O with energy in regions I and II when the shock moves to 3.5 $R_{\odot}$, and region III when the shock moves to 3 $R_{\odot}$ and 3.5 $R_{\odot}$.
The values in each curve have been normalized to that at 100 keV/nuc.
}
\label{fig:Fe2O}
\end{figure}

Figures \ref{fig:Fe2O}(a)-(b) show the spatial distributions of Fe and O with energies $>$10 MeV/nuc when the shock moves to 3.5 $R_{\odot}$ for the case of turbulence spectral index $\Gamma = 1.1$. The intensity of O is much higher than that of Fe, indicating a more efficient acceleration of O.
Same as in Figure \ref{fig:double}, we select three localized regions along the shock front, nonstreamer region (I), streamer region (II), and the transition region (III).
Figures \ref{fig:Fe2O}(c)-(d) plot the variations of Fe/O with energy. Since we only focus on the variability of energy-dependence of Fe/O rather the realistic values, we have normalized the curves to the corresponding value at 100 keV/nuc.
At low energies, Fe/O in the three regions generally remains constant.
This is because for different ion species the particle spectral shapes below the break energy are essentially the same, as predicted by the DSA theory (see Figure \ref{fig:different}(c)).
However, at higher energies, the energy-dependence of Fe/O shows significantly different trends.
In regions I and II, Fe/O drops promptly with energy above $\sim$10 and $\sim$40 MeV/nuc, respectively.
In region III, Fe/O first drops with energy between $\sim$10-50 MeV/nuc, then it increases with energy between $\sim$50-150 MeV/nuc, and then it decreases again.
Based on our previous analysis, we suggest that this is due to the contribution of high-energy particles escaping from the streamer region (II) where Fe/O is much larger than that of locally accelerated particles.
Note that the curve when the shock moves to 3 $R_{\odot}$ shows that Fe/O at $\sim$150-200 MeV/nuc is even larger than the value below 10 MeV/nuc.  Since Fe particles have a larger diffusion coefficient than O at the same energy/nuc, they transport from the streamer region to other regions faster, leading to this enhanced Fe/O. This feature may be detected if an observer is connected magnetically with region III.
We also find similar energy dependence for other abundance ratio, e.g., He/H.

\section{Conclusions and Discussion} \label{sec:Conclusions and Discussion}
In this work, we numerically model the acceleration of both protons and heavier ions at coronal shocks and investigate the double-power-law feature in the particle energy spectra.
We find that the energy spectra of all ion species integrated over the whole simulation domain are approximately double power-laws and the break energy $E_B$ has a power-law dependence on the ion charge-to-mass ratio as $E_B \sim (Q/A)^{\alpha}$, with $\alpha$ varying between 0.16$-$1.2.
The correlation of $\alpha$ with the turbulence spectral index $\Gamma$ agrees well with the theoretical predication, $\alpha = 2(2-\Gamma)/(3-\Gamma)$, inferred from the ``equal diffusion coefficient'' condition \citep{cohen05}.
By considering enhanced turbulence in the form of self-generated Alfv$\acute{e}$n waves, \citet{li09} showed that $\alpha$ can reach 2 for a quasi-parallel shock.

We also examine the particle spectra in three localized regions close to the Sun. We find that in the transition region between the streamer and nonstreamer regions the particle spectra can be well described by a double power law, while in the localized streamer and nonstreamer regions the particle spectra exhibit a power-law with exponential rollover. We suggest that the superposition scenario, as shown in Figure 1, can explain the formation of the double power law distribution in this transition region. $>$100 MeV/nuc high-energy particles mainly originate from the streamer region where particles are more efficiently accelerated, and the compound populations of higher-energy particles diffused from the streamer region and locally accelerated relatively lower-energy particles can produce a double-power-law spectral shape. Although at 1 AU, these local spectral features may have been washed out due to further mixing effects during SEP propagation, the upcoming Parker Solar Probe and Solar Orbiter observations may provide critical tests to this prediction. In addition, we suggest that the diffusion and mixing of energetic particles can also provide an explanation for the puzzling problem of the variability in the energy dependence of Fe/O.
The variations of Fe/O with energy in the transition region are similar to the simulation results in \citet{sandroos07}, however, they included flare suprathermals as the seed population.

We note that the DSA theory applies when the anisotropy is small, leading to the so-called “injection problem”. The injection energy has a dependence on the shock normal angle and is larger for quasi-perpendicular shocks \citep{zank06,giacalone17}. Hybrid simulations have shown that even thermal particles can be efficiently accelerated at perpendicular shocks if large-scale turbulence is present \citep{giacalone05a,giacalone05b}. We find that when we consider the dependence of injection energy on the shock geometry, the double power law distribution may still be obtained and our proposed superposition scenario remains reasonable.
In addition, the number density in the corona varies from the streamer to the open field region and may affect the particle injection rate along the shock front. The coronal (electron) density can be deduced from coronagraph polarized brightness (pB) images and it is found that the density in the streamer is about a few times of the ambient corona \citep[e.g.,][]{chen11,kwon13}. We find that a double power law can still emerge when we set the ratio of particle injection rate between the streamer and open field region to be 10. This is because only a small fraction of high-energy particles diffused from the streamer contribute to the high-energy portion of the double power law in region III (about 5\% as shown in Figure 4(e)).

Our simulations reveal that the double power law distribution can occur near the Sun as a result of particle acceleration at CME-shocks. It also predicts that the spectral forms near the Sun can vary significantly along the shock front and depend on the region where the observer is connected.
In observations, previous studies have shown that if one SEP event is detected by multiple spacecraft from different points of view, the energy spectral shapes differ largely from one spacecraft to another for the same event \citep[e.g.,][]{cohen17,mewaldt13}.
Note that the double power laws as observed in large SEP events are event-integrated fluence spectra measured at 1 AU. The spectral properties of energetic particles may be modified during interplanetary transport due to effects including adiabatic deceleration and scattering \citep{li15,zhao16}. However, \citet{mason12} showed that although the break energy slightly decreases due to transport effects, the basic spectral form remains unchanged, therefore spectral breaks should be present near the Sun.
Recently launched Parker Solar Probe and Solar Orbiter can provide SEP measurements near the Sun, where the acceleration takes place and transport effects are reduced, and will advance our understanding of spectral properties in SEP events.

\acknowledgments
This work was supported by the National Natural Science Foundation of China under grants 42074203, 11873036 and 11790303 (11790300), the Young Elite Scientists Sponsorship Program by China Association for Science and Technology, and the Young Scholars Program of Shandong University, Weihai.
F. G. acknowledges support from DOE grant DE-SC0020219 and NSF grant 2109154, and the LDRD program at Los Alamos National Laboratory.



\appendix

\section{Derivation of the Particle Diffusion Coefficient from the Quasi-linear Theory} \label{sec:Appendix A}
Following the approach of \citet{giacalone99}, we derive the particle diffusion coefficient from the quasi-linear theory \citep{jokipii66}.

The turbulence power spectrum $P(k)$ is in the form of
\begin{equation}
P(k) = A_0 L_c \sigma^2 B_0^2 \frac{1}{1 + (k L_c)^{\Gamma}},
\end{equation}

\noindent where $k$ is the wave number, $L_c$ is the turbulence correlation length, $\Gamma$ is the spectral index in the inertial range, $\sigma^2$ is the variance of turbulent magnetic field, $A_0$ is the normalization constant, such that the integral over $k$ gives the turbulence level,
    \begin{equation}
    \int_0^\infty P(k) dk = \sigma^2 B_0^2 = \langle \delta B \rangle ^2.
    \end{equation}
    
In Equation (A1), $A_0=[\int_{k_{min}}^{k_{max}}\frac{d(kL_c)}{1+(kL_c)^\Gamma}]^{-1}$, $k_{min}$ and $k_{max}$ correspond to the largest and smallest wavelengths in the system. Here the integral is taken from 0 to $\infty$ for $\Gamma > 1$, so it can be easily obtained.

In the standard quasi-linear theory, the resonant interaction between the particle and the turbulent magnetic field can be related by the pitch-angle diffusion coefficient $D_{\mu \mu}$ \citep{jokipii71},
\begin{equation}
D_{\mu \mu} = \frac{\pi}{4} \Omega_0 (1-\mu^2) \frac{k_r P(k_r)}{B_0^2},
\end{equation}
where $k_r=|\Omega_0/v\mu|$ is the resonant wavenumber, $\Omega_0$ is the particle gyrofrequency, $\mu$ is the pitch angle cosine.

The diffusion coefficient along the parallel direction of the magnetic field can be related to the pitch angle diffusion \citep{jokipii66,earl74,luhmann76} by
\begin{equation}
\kappa_\|(v) = \frac{v^2}{4} \int_0^1 \frac{(1 - \mu^2)^2}{D_{\mu \mu}} d\mu,
\end{equation}
where $v$ is particle's velocity.

Substituting Equations (A1) and (A3) into Equation (A4), we get
    \begin{equation}
    \kappa_\parallel= \frac{v^3}{A_0 \pi L_c \Omega_0^2 \sigma^2}
    \left[\frac{1}{4}+\left(\frac{\Omega_0L_c}{v}\right)^{\Gamma}\frac{2}{\left(2-\Gamma\right)
    \left(4-\Gamma\right)}\right].
    \end{equation}
    
For the Kolmogorov spectrum with $\Gamma$ = 5/3, the equation will be equivalent to that given in \citet{giacalone99} and used in our previous studies \citep{kong17,kong19}. If we take $B$ = 1 G, $L_c$ = 0.01 $R_\odot$, $\sigma^2=\langle \delta B \rangle^2/B_0^2=0.14$, we can get $\kappa_{\parallel0} = 1.4\times10^{17}$cm$^2$ s$^{-1}$ for protons of $E_0 =$ 100 keV.

In this study, we consider $\Gamma \geqslant$ 0.5, and particle energy $<$1 GeV/nuc, so $\left(\frac{\Omega_0 L_c}{v}\right)^{\Gamma} \frac{2}{\left(2-\Gamma\right)\left(4-\Gamma\right)} > \frac{1}{4}$, 
and Equation (A5) can be simplified as
\begin{equation}
\kappa_\|(v) = \frac{2}{(2 - \Gamma)(4 - \Gamma)} \\
\frac{L_c}{A_0 \pi \sigma^2} \\
\left( \frac{e B L_c}{m_p} \right)^{\Gamma - 2} \\
\frac{v^{3 - \Gamma}}{\gamma^{\Gamma - 2}} \\
\left( \frac{Q}{A} \right)^{\Gamma - 2},
\end{equation}
where $\gamma$ is the Lorentz factor, $m_p$ is the proton mass, $e$ is the elementary charge, and for different ion species the gyrofrequency $\Omega = (Q/A)(eB/\gamma m_p)$.

We assume the ions have the same initial velocity as protons, i.e., $E_{0,i} =$ 100 keV/nuc. Particle momentum $p = m v = \gamma A m_p v$,
and $p_0 = \gamma_0 A m_p v_0$ corresponding to the injection energy $E_0$ with $\gamma_0 = 1.0001 \approx 1$.
So Equation (A6) can be written as
    \begin{equation}
    \kappa_{\parallel}(p) = \kappa_{{\parallel 0}} \left(\frac{p}{p_0}\right)^{3 - \Gamma} \left(\frac{Q}{A}\right)^{\Gamma - 2} \frac{\gamma_0}{\gamma}\\,
    \end{equation}
where 
\begin{equation}
\kappa_{\parallel0} = \frac{2}{(2 - \Gamma)(4 - \Gamma)} \\
\frac{L_c}{A_0 \pi \sigma^2} \\
\left( \frac{e B L_c}{m_p} \right)^{\Gamma - 2} \\
\frac{v_0^{3 - \Gamma}}{\gamma_0^{\Gamma - 2}} \\.
\end{equation}

For the same form of turbulence power spectrum $P(k)$, $\kappa_{\parallel0}$ remains a constant for different ions.
In this work, we also examine the effects of turbulence spectra by considering different spectral indices. 

Note that the form of diffusion coefficient in Equation (A7) is similar to Equation (10) as derived in \citet{li13}, except that they only consider electrons with $Q/A$ = 1.



\end{document}